\documentclass[prl,twocolumn,twoside,preprintnumbers,superscriptaddress,nofootinbib]{revtex4}

\usepackage{color}
\usepackage{bbold}
\usepackage{mathrsfs}
\usepackage{amsmath}
\usepackage{graphicx}
\usepackage{dcolumn}
\usepackage{bm}
\usepackage[hyperfootnotes=false]{hyperref}
\usepackage{xspace}
\usepackage{amsmath, amsthm, amssymb, mathtools,empheq,latexsym,dsfont}

\newcommand{\be}{\begin{equation}}
\newcommand{\ee}{\end{equation}}

\newcommand{\bea}{\begin{eqnarray}}
\newcommand{\eea}{\end{eqnarray}}

\newcommand{\ba} {\begin{eqnarray}}
\newcommand{\ea} {\end{eqnarray}}
\newcommand{\dd} {\displaystyle}

\def\article{\@ifnextchar[{\earticle}{\oarticle}}

\begin{document}
\title{The irresistible call of $\tau=i$}
\author{Ferruccio Feruglio}
\affiliation{Istituto Nazionale di Fisica Nucleare, Sezione di Padova, I--35131 Padova, Italy}

\begin{abstract}
We analyze a large set of modular invariant models of lepton masses and mixing angles, pointing out that
many of them prefer to live close to the self-dual point $\tau=i$. We show that in the vicinity of this point
a universal behavior naturally emerges, independently from details of the theory such as the finite modular group
acting on the lepton multiplets, the weights of the matter multiplets and even the form of the kinetic terms, which are not required to be neither minimal nor flavour blind. The neutrino mass spectrum is 
normally ordered and universal relations describe the scaling of the physical observables in terms of the
parameter $|\tau-i|$.
\end{abstract}
\maketitle
%
\noindent 
{\bf Introduction}

\noindent
Of all unsolved problems in particle physics, the flavor puzzle is perhaps the most intriguing one. Despite the rich and precise available data, a baseline theoretical interpretation of them
is still missing. Most attempts to characterize the fermion spectrum are based on the idea of flavor symmetry,
an alleged invariance of the theory under transformations among members of different generations.

Exact flavor symmetries are not realized in nature and a symmetry breaking mechanism is required, representing
the source of most complications and loss of predictability in realistic constructions. 
One of the simplest order parameter of flavor symmetry breaking is provided
by the modulus $\tau$, a single complex field living in the upper part of the complex plane.
The modulus is a key degree of freedom in string theory where it describes the geometry of two extra dimensions arising from the partial compactification of the ten dimensional spacetime. An intrinsic redundancy in this description is removed by a set of discrete modular transformations \cite{Hamidi:1986vh,Dixon:1986qv,Lauer:1989ax,Lauer:1990tm}, playing the role of gauge symmetries.
Such a setup offers a simple and well-motivated framework to the idea of flavor symmetry \cite{Baur:2019kwi}.
A finite copy of the modular group acts in generation space. At the same time the
modulus transforms non trivially under the modular group and plays the role of symmetry breaking order parameter. Moreover, this setup is suitable to be explored in a bottom-up approach~\cite{Feruglio:2017spp}. 

Despite its simplicity and appeal, the approach is undermined by two major factors. On the one hand,
there is a large freedom in choosing the transformation properties of the matter fields. Unlike the more familiar linear transformations, the modular ones are specified by two additional integers, the weight and the level.
Varying these extra parameters, hundreds of explicit realizations have been 
proposed in the last years. On the other hand, modular invariance proved too weak in 
constraining the kinetic terms of the matter multiplets \cite{Chen:2019ewa}. Such a drawback is usually bypassed by assuming flavor-universal kinetic terms, but 
this assumption cannot be justified in a bottom-up approach, which allows for more general and
flavor-dependent kinetic terms, thus reducing the predictability.

The variety of different realizations allowed in a pure bottom-up approach have not permitted so far the identification of a unique scenario. What can be learned from the large number of models already proposed?
Is there any common feature emerging from the different constructions?

In this letter we analyze the existing modular invariant flavor models describing the lepton sector. 
Most of them successfully reproduce the data in terms of a limited number of free parameters,
which include the modulus $\tau$ itself, varied to maximize the agreement between data and theory.
We start by observing that, in a large number of cases, the preferred value of the modulus falls close
to the self-dual point $\tau=i$, where the modular transformation $\tau\to-1/\tau$ is unbroken. This preference is
even more pronounced for the subclass of CP invariant models.
We investigate the behavior of these models in a neighborhood of $\tau=i$, also allowing for the most general kinetic terms. We find that, quite surprisingly, all these models exhibit a universal behavior, which does not depend 
on the details of the theory, such as its level, the modular weights of matter fieds and the precise form of the kinetic terms. 

This class of models predicts normal ordering of
neutrino masses. Small quantities such as $\Delta m^2_{sol}/\Delta m^2_{atm}$,
$\sin^2\theta_{13}$ and $(\sin^2\theta_{12}-1/2)$ have well-defined scaling laws in terms of the parameter $|u|=|(\tau-i)/(\tau+i)|$.
All seemingly different and independent modular invariant lepton models, built by varying
all possible finite modular symmetries, all weights of the matter fields, and adopting the most general kinetic terms, 
fall in the same universality class when $\tau$ approaches the self-dual point. 
\vspace{5pt}

\noindent
{\bf Modular invariant models of lepton masses}
%

\begin{figure}[h!]
\centering
\includegraphics[width=0.5\linewidth]{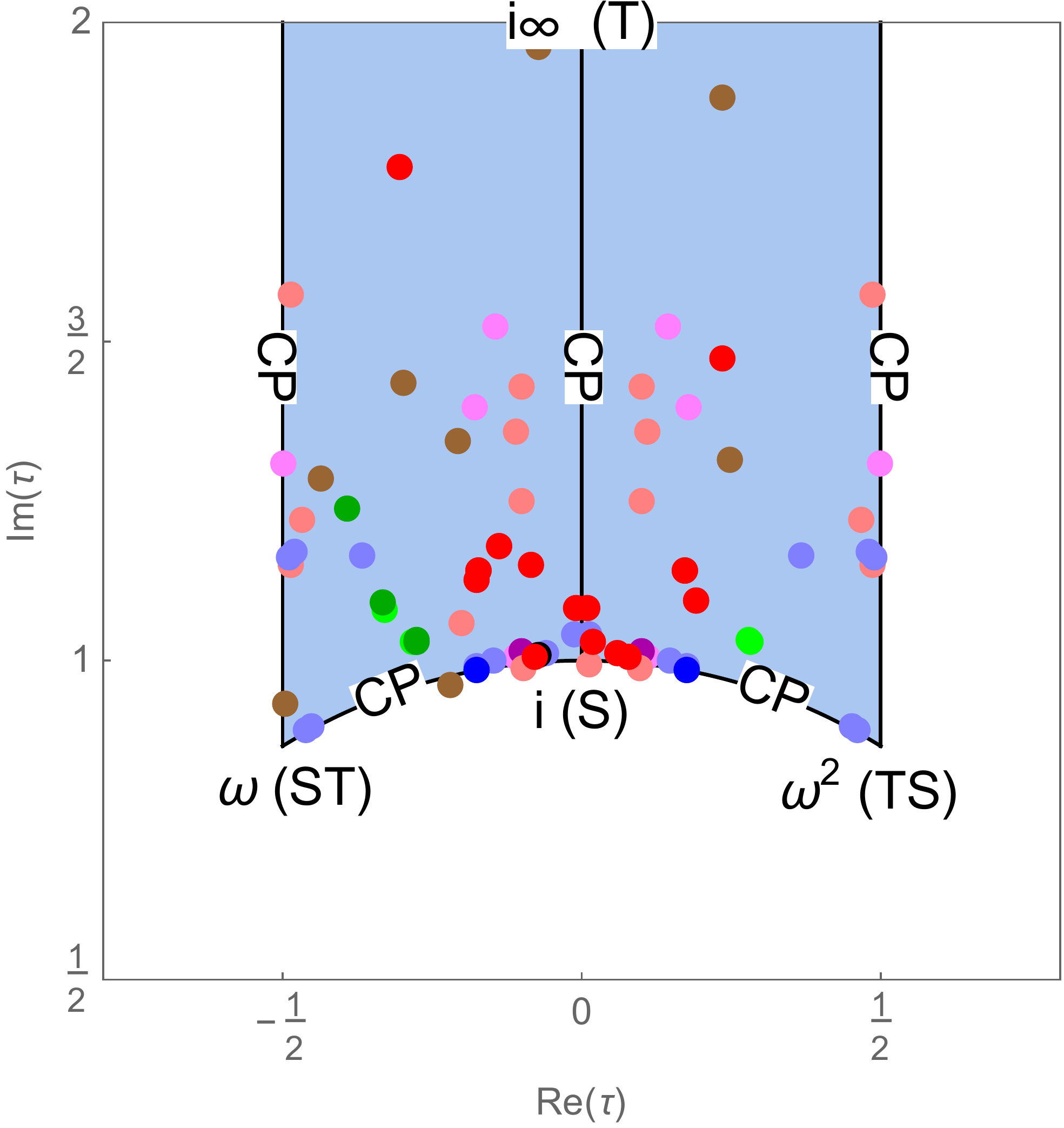}\includegraphics[width=0.5\linewidth]{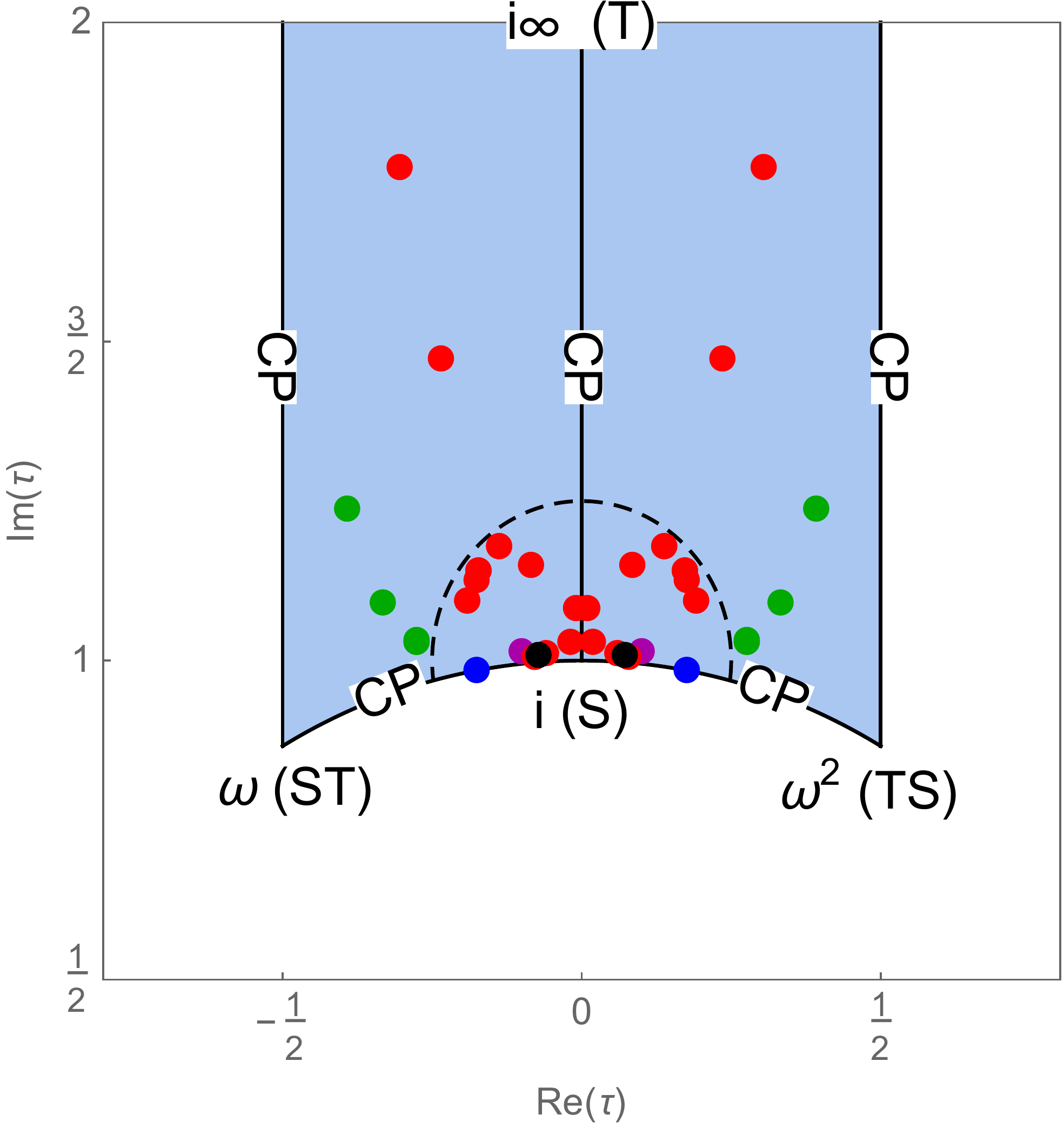}
\caption{Fundamental domain ${\cal F}$ (light blue region) and fixed points (see text). Dots are the best fit values of $\tau$ in models of ref. \cite{Criado:2018thu,Ding:2019zxk} ($\Gamma_3$ - light red), 
\cite{Yao:2020qyy,Okada:2020brs} ($\Gamma_3 \& CP$ - red),
\cite{Novichkov:2018ovf} ($\Gamma_4$ - light magenta), 
\cite{Novichkov:2019sqv} ($\Gamma_4 \& CP$ - magenta), 
\cite{Liu:2020akv} ($\Gamma_4'$ - light blue), 
\cite{Liu:2020akv} ($\Gamma_4' \& CP$ - blue),
 \cite{Wang:2021mkw} ($\Gamma_5' \& CP$ - black), 
 \cite{Li:2021buv} ($\Gamma_6'$ - light green), 
 \cite{Li:2021buv} ($\Gamma_6' \& CP$ - green), 
 \cite{Ding:2020msi} ($\Gamma_7$ - brown). 
 We use the notation  $\Gamma_N'=SL(2,\mathbb{Z}_N)$ and  $\Gamma_N=SL(2,\mathbb{Z}_N)/\{\pm \mathbb{1}\}$.
 In the left panel all models are displayed. The right panel includes only CP invariant models, for which the full pair of points
 $\tau$ and $-\bar\tau$ is shown. The dashed line
 represents the contour $|\tau-i|=0.25$.}
\label{lines}
\end{figure}

\noindent
We focus on (${\cal N}=1$ rigid) supersymmetric modular invariant theories whose field content includes, among the chiral supermultiplets, the modulus $\tau$, whose scalar component has a positive definite imaginary part, and a set of matter fields $\varphi$. The relevant part of the action reads:
\be
\label{one}
{\mathscr S}=\int d^4 x d^2\theta d^2\bar\theta~ K+\int d^4 x d^2\theta~ w+\int d^4 x d^2\bar\theta~ \bar w~,
\ee
where $K$ is the K\"ahler potential, describing kinetic terms and $w$ is the superpotential, describing Yukawa interactions. Supersymmetry breaking effects have been shown to be negligible in a large portion of the parameter space~\cite{Criado:2018thu}, especially for normally ordered neutrinos, and will be ignored here.
Modular invariance requires ${\mathscr S}$ to remain unchanged under the transformations:
\begin{align}
\label{two}
\begin{array}{l}
\tau\xrightarrow{\gamma}\dd\frac{a\tau+b}{c\tau+d}~~~~~~~~~~~~~
\gamma=
\left(
\begin{array}{cc}
a&b\\
c&d
\end{array}
\right)\\[10 pt]
\varphi \xrightarrow{\gamma} (c\tau+d)^{-k_\varphi}\rho^N_\varphi(\gamma)\varphi~~~.
\end{array}
\end{align}
The matrix $\gamma$ belongs to the modular group $SL(2,\mathbb{Z})$ ($a$, $b$, $c$ and $d$ are integer and $ad-bc=1$), generated by the elements
\be
S=\left(
\begin{array}{cc}
0&1\\-1&0
\end{array}
\right)~,~~~~~~~~~~
T=\left(
\begin{array}{cc}
1&1\\0&1
\end{array}
\right)~.
\ee
The transformation law of $\varphi$ is characterized by a unitary
representation $\rho^N_\varphi$ of the finite modular group $SL(2,\mathbb{Z}_N)$ and by the integer weight $k_\varphi$ 
(irreducible components of $\varphi$ admitting independent weights). The choice of $N$, the level of the representation common to all matter fields, 
specifies the finite flavour group effectively acting on $\varphi$. For the first few levels $N=2,3,4,5$ the group $SL(2,\mathbb{Z}_N)$ is isomorphic to the double covering of small permutation groups $S_3,A_4,S_4,A_5$, respectively.
The modular group allows to restrict the scalar component of the modulus within the fundamental domain ${\cal F}=\{{\tt Im}\tau>0, |{\tt Re}\tau|\le 1/2, |\tau|\ge 1\}$. In a generic point of ${\cal F}$ the modular group is
fully broken. At the fixed points, $\tau=(i,\omega\equiv-1/2+i\sqrt{3}/2,i \infty)$ a finite subgroup of $SL(2,\mathbb{Z})$ is preserved. Invariance under CP requires ${\mathscr S}$ to remain unmodified under the transformations (we use a bar(asterisk) to denote conjugation of fields(numbers))~\cite{Baur:2019kwi,Novichkov:2019sqv}:
 \begin{align}
\tau\xrightarrow{CP}-\bar\tau~~~~~~~~~~~~~
\varphi \xrightarrow{CP} \bar\varphi~~~.
\end{align} 
If the theory is CP invariant, CP violation can only arise from a spontaneous breaking. This occurs everywhere in ${\cal F}$,
except along the imaginary $\tau$ axis and along the border of ${\cal F}$, see fig. \ref{lines}.

Such a setup has been exploited to describe lepton masses, mixing angles and phases in a large number of models.
The matter multiplets $\varphi$ include two Higgs doublets $H_{u,d}$, electroweak doublets $L$, singlets $E^c$ and, when neutrino masses arise from the seesaw mechanism, heavy Majorana singlets $N$. To minimize the number of free parameters, $L$ is usually assigned
to an irreducible triplet of $SL(2,\mathbb{Z}_N)$. We have investigated more than 100 such models that reproduce accurately the data and predict nine physical quantities (neutrino masses, mixing angles and phases) in terms of the value of $\tau$ and  2 or 3 additional Lagrangian parameters. They adopt flavour universal kinetic terms, but rely on different choices of level and weights. 
In all cases, $\tau$ is treated as a free parameter, scanned to maximize the agreement between data and theory. There is no prejudice about the value of $\tau$, nor about possible dynamical mechanisms that can determine $\tau$ or favour some region of ${\cal F}$. In fig. \ref{lines} (left panel) we plot 103 best fit points. 

Models are not equally spread in ${\cal F}$. There is a preference
for regions along the boundary of ${\cal F}$, and in particular around the fixed point $\tau=i$.
If we focus on models where CP is spontaneously broken (right panel of fig. \ref{lines}) such a preference is more pronounced. CP invariant models evaluated at symmetric points $\tau$ and $-\bar\tau$
only differ in the sign of the CP violating phases, and have been considered equally successful in fig. \ref{lines} (right panel) that includes $27$ pairs of points. Two thirds of these points fall close to the self-dual point: $|\tau-i|<0.25$.
We regard this feature as an indication of an intrinsic property of the theory, deserving an explanation.
\begin{figure}[h!]
\centering
\includegraphics[width=0.75\linewidth]{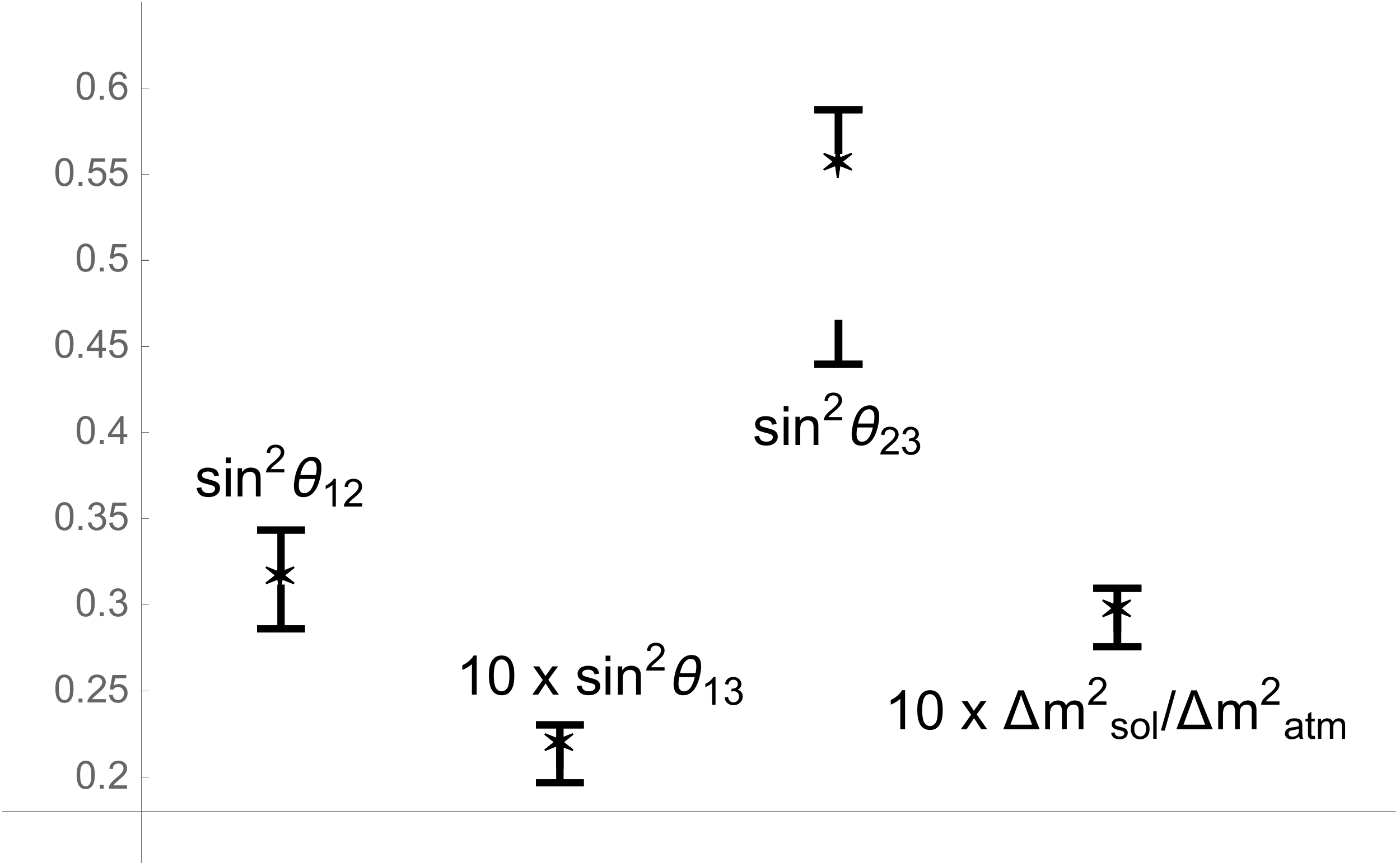}\\[15 pt]
\includegraphics[width=0.75\linewidth]{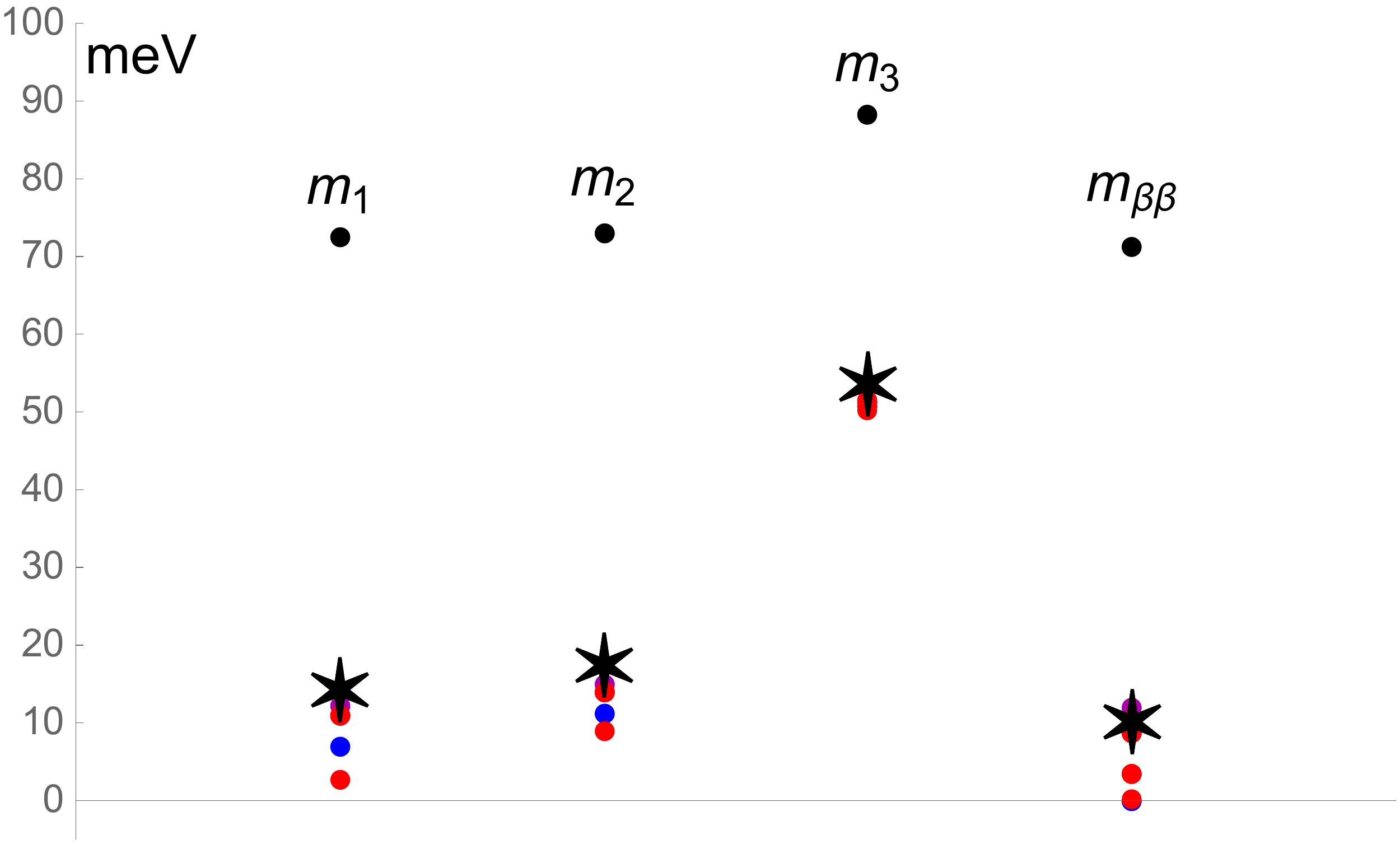}\\[15 pt]
\includegraphics[width=0.75\linewidth]{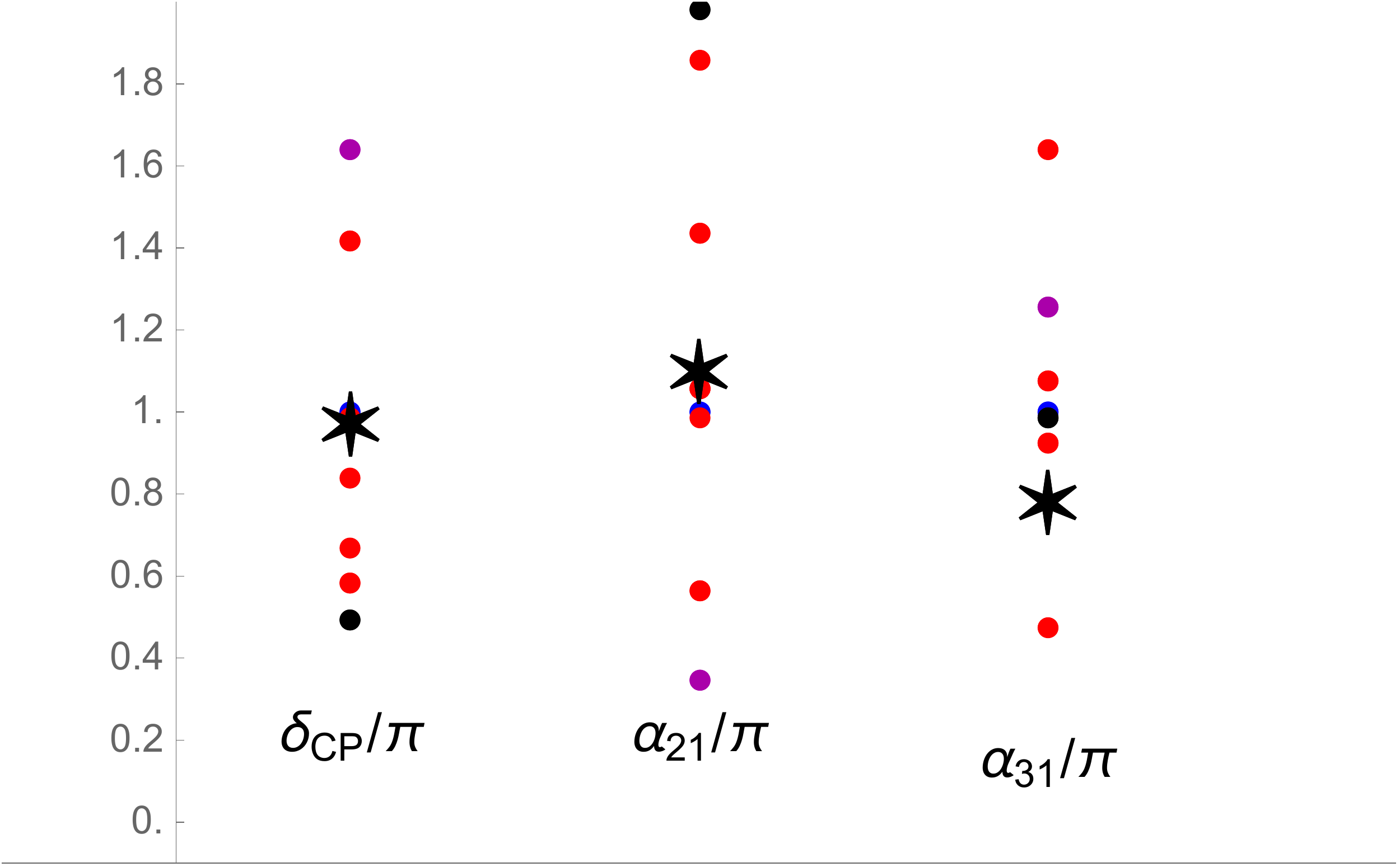}
\caption{Mixing angles, mass parameters and phases of 14 pairs of CP and modular invariant models featuring normal ordering and $|\tau-i|<0.25$ (see text). For mixing angles and $\Delta m^2_{sol}/\Delta m^2_{atm}$, the intervals covered by the model predictions are shown (upper panel). For neutrino masses, $m_{\beta\beta}$ and CP violating phases, the full distributions of predictions
are displayed (lower panels). The color code is identical to the one of fig. \ref{lines}. A star shows the average over the 14 models. CP violating phases refer to models where ${\tt Re}\tau>0$.}
\label{angles and masses}
\end{figure}

In fig. \ref{angles and masses} we show the predictions of a homogeneous set of $14$ pairs of CP invariant models. We have discarded nine pairs by requiring $|\tau-i|<0.25$, two pairs not relying on the seesaw mechanism, one pair - the only one - predicting inverted ordering and a last pair, for which full data were not available. The average of some combinations of observables over $13$ models of our sample is listed in eq. (\ref{ave})~\footnote{We have removed the model of ref. \cite{Wang:2021mkw}, denoted by black points in the fig. \ref{angles and masses}, for which $\rho_L$ is reducible.}.
The quoted error has no statistical meaning: it only illustrates the spread of the actual predictions for each combination.
All  the dimensionless values are of the same order of magnitude and close to the average deviation of $\tau$ from the imaginary unit.
We will see that this behavior is a universal feature of modular invariant models living close to  $\tau=i$.
\begin{align}
\label{ave}
\begin{array}{lcl}
\vert \tau-i\vert=0.20\pm0.04&&|u|=|\frac{\tau-i}{\tau+i}|=0.095\pm 0.015\\[10 pt]
\sum_i m_i=73.9\pm 4.6~{\rm meV}&&m_{\beta\beta}=5.5\pm4.2~{\rm meV}\\[10 pt]
\frac{m_2+m_1}{2}=11.5\pm 2.2~{\rm meV}&&
\frac{m_2+m_1}{2 m_3}=0.23\pm 0.04\\[10 pt]
2\frac{m_2-m_1}{m_2+m_1}=0.34\pm 0.25&&
\left(\frac{\Delta m^2_{sol}}{\Delta m^2_{atm}}\right)^{1/3}=0.310\pm 0.001\\[10 pt]
\sin\theta_{13}=0.148\pm 0.001&&
\left\vert \dd\frac{1}{\sqrt{2}}-\sin\theta_{12}   \right\vert=0.14\pm 0.01~.
\end{array}
\end{align}
\vspace{20pt}

\noindent
{\bf Universal behavior near $\tau=i$}

\noindent
At $\tau=i$ both CP and the $\mathbb{Z}_4$ subgroup of $SL(2,\mathbb{Z})$
generated by the element $S$, are preserved. The element $S$ acts non linearly in the field space and it is convenient to move to a basis where CP and $\mathbb{Z}_4$ are linearly realized. We define \cite{Feruglio:2021dte,Novichkov:2021evw}:
\be
u=\dd\frac{\tau-i}{\tau+i}~,~~~~~~~~~~\Phi=(1-u)^{k_\varphi}\varphi~.
\ee
The new fields transform as:
\begin{align}
\label{linear}
\begin{array}{lcccl}
u\xrightarrow{S} -u&&&&\Phi\xrightarrow{S}\Omega_\varphi(S)~\Phi\\[10pt]
u\xrightarrow{CP} \bar u&&&&\Phi\xrightarrow{CP} \bar\Phi~,
\end{array}
\end{align}
where the unitary matrices $\Omega_\varphi(S)$ are
\begin{align}
\Omega_\varphi(S)= i^{k_\varphi} \rho_\varphi(S)~.
\end{align}
The most general (${\cal N}=1$, rigid) supersymmetric theory depending on $u$ and $\Phi$ and invariant under the linear transformations (\ref{linear}), extends the setup of eqs. (\ref{one}) and (\ref{two}). Indeed ${\mathscr S}$ is also invariant under the transformations generated by the element $T$ which, evaluated in terms of $u$, reads:
\begin{align}
\label{tu}
u\xrightarrow{T} \dd\frac{1-(1-2i) u}{(1+2i)-u}=\frac{1-2i}{5}+\frac{4}{25}(3+4 i)u+...~~~.
\end{align}
The new, more general, theory has non-canonical and flavour-dependent kinetic terms.
After moving to the basis where such terms are canonical, it is not difficult to prove that
charged lepton and neutrino mass matrices, $m_e^2(u,\bar u)\equiv m_e(u,\bar u)^\dagger m_e(u,\bar u)$ and $m_\nu(u,\bar u)$, should satisfy the relations:
\begin{align}
\label{CP3}
m_\nu(\bar u,u)=m_\nu(u,\bar u)^*~,~~m_e^2(\bar u,u)=[m_e^2(u,\bar u)]^*~,
\end{align}
and
\begin{align}
\label{trans1}
\begin{array}{l}
\Omega(S)^T m_\nu(-u,-\bar u)~ \Omega(S)=m_\nu(u,\bar u)\\[5pt]
\Omega(S)^\dagger m_\nu(-u,-\bar u)^{-1}~ \Omega(S)^*=m_\nu(u,\bar u)^{-1}\\[5pt]
\Omega(S)^\dagger~ m_e^2(-u,-\bar u)\Omega(S)=m_e^2(u,\bar u)~,
\end{array}
\end{align}
where we have defined:
\begin{align}
\label{trans2}
\Omega(S)\equiv \Omega_{H_u}(S)\Omega_L(S)=i^{k_{H_u}+k_L} \rho_{H_u}(S)\rho_L(S)~.
\end{align}
A key observation of the present note is that, under the only assumption that $\rho_L$ is irreducible,
which is adopted in the vast majority of existing models, the combination 
$\Omega(S)$ is completely fixed up to an overall phase factor. Once diagonalized, $\Omega(S)$ reads:
\begin{align}
\label{Omega}
\Omega(S)=i^{k_S}{\tt diag}(1,-1,-1)~,
\end{align} 
where $k_S$ is an integer depending on $k_{H_u}$, $k_L$ and $\rho_{H_u}$, as well as on the specific
3-dimensional representation $\rho_L$ of $SL(2,\mathbb{Z}_N)$. This result
can be proved by directly inspecting all 3-dimensional irreducible representations
of $SL(2,\mathbb{Z}_N)$ that, for $N$ a power of 2 or a prime, exist only for $N=3,4,5,7,8,16$~\cite{nobs}. There are 33 inequivalent such representations, easily rebuildable from Appendix A of ref. \cite{Eholzer:1994th}. 
For these levels we can also construct the 1-dimensional representations of $SL(2,\mathbb{Z}_N)$. A straightforward computation leads to eq. (\ref{Omega}), valid for all $N$. It is also possible to prove eq. (\ref{Omega}) without relying on explicit representations~\footnote{G.-J. Ding, private communication.}.

Working in the vicinity of $u=0$ and exploiting eq. (\ref{Omega}), we can explicitly write the expansion of  $m_e^2(u,\bar u)$ and $m_\nu(u,\bar u)$ in powers of $u$ and $\bar u$. 
We cannot make us of the invariance under $T$, since
the transformation of eq. (\ref{tu}) spoils the power expansion.
Here we discuss the case of $k_S$ odd and $m_\nu(0,0)$ singular,
relevant if the neutrino masses arise from the seesaw mechanism.
After moving to the basis where $m_e^2(u,\bar u)$ is diagonal, we obtain:
\begin{align}
\label{caseb}
m_\nu^{-1}=m_{0\nu}^{-1}
\left(
\begin{array}{lll}
x_{11}~ x&x^0_{12}&x^0_{13}\\
\cdot&x_{22}~ x&x_{23}~ x\\
\cdot&\cdot&x_{33}~ x\\
\end{array}
\right)+{\cal O}(x^2)~,
\end{align} 
where $m_{0\nu}$ is a mass parameter and we used the definitions
\begin{align} 
\begin{array}{l} 
u=x~ e^{\dd i \theta}~~~~~~~~~(x>0,2 \pi> \theta\ge 0)\\[8pt]
x_{ij}~ x\equiv(x^{10}_{ij}e^{\dd i \theta}+x^{01}_{ij}e^{\dd -i \theta})~x ~,
\end{array}
\end{align}
$x^0_{12}$, $x^0_{13}$, $x^{10}_{ij}$ and $x^{01}_{ij}$ being real coefficients.
From eq. (\ref{caseb}) we derive the first few terms in the $x$-expansion of neutrino masses, mixing angles and phases:
\begin{align}  
\label{expansions}
\begin{array}{l}
m_{1,2}=\dd\frac{m_{0\nu}}{h}\left(1\mp\dd\frac{s}{2h}x\right)~~~~~~~~~~~
m_3=\dd\frac{m_{0\nu}}{|q| x}\\[10 pt]
\sin^2\theta_{12}=\frac{1}{2}\left(1+\dd\frac{l\bar k+\bar l k}{h s}x\right)~~~~~
\sin^2\theta_{13}=2 \dd\frac{|n|^2}{h^2} x^2\\[10 pt]
\sin^2\theta_{23}=\dd\frac{(x^0_{13})^2}{h^2}(1+{\cal O}(x))\\[10 pt]
\delta_{CP}=\pi-\arg\left[\dd\frac{(c_\nu-is_\nu)^2 x^0_{12} x^0_{13}}{n}\right]+{\cal O}(x^2)\\[10 pt]
\alpha_{21}=\pi+{\cal O}(x)\\[10 pt]
\alpha_{31}=\arg(q)-\arg\left[(c_\nu-is_\nu)^2\right]+{\cal O}(x)~.
\end{array}
\end{align}
We have defined:
\begin{align}  
\label{defs}
\begin{array}{l}
h=\sqrt{(x^0_{12})^2+(x^0_{13})^2}~~~~~~~~~s=\sqrt{(k+\bar k)^2-(l-\bar l)^2}\\[7 pt]
k=+\dd\frac{x_{11}}{2}+\dd\frac{1}{2h^2}
\left((x^0_{12})^2 x_{22}+(x^0_{13})^2 x_{33}+2 x^0_{12} x^0_{13} x_{23}\right)\\[7 pt]
l=-\dd\frac{x_{11}}{2}+\dd\frac{1}{2h^2}
\left((x^0_{12})^2 x_{22}+(x^0_{13})^2 x_{33}+2 x^0_{12} x^0_{13} x_{23}\right)\\[7 pt]
n=\dd\frac{1}{2h^2}\left[\left(x^0_{12})^2 - (x^0_{13})^2 \right) x_{23} + x^0_{12} x^0_{13} \left(-x_{22} + 
x_{33}\right)\right]\\[7 pt]
q=\dd\frac{1}{2h^2}\left( (x^0_{13})^2 x_{22}+(x^0_{12})^2 x_{33}-2 x^0_{12} x^0_{13} x_{23}\right)\\[7 pt]
\dd\frac{2 c_\nu s_\nu}{c_\nu^2- s_\nu^2}=i\dd\frac{l-\bar l}{k+\bar k}+{\cal O}(x^2)~~~~~~~~~i(l-\bar l)={\mathcal O}(1)~.
\end{array}
\end{align}
Other combinations of interest are:
\begin{align}  
\label{ofinterest}
\begin{array}{l}
\dd\frac{m_1+m_2}{2 m_3}=\frac{|q|}{h}x~~~~~~~~~~~
2\dd\frac{m_2-m_1}{m_2+m_1}=\frac{s}{h}x\\[10 pt]
\left(\dd\frac{\Delta m^2_{sol}}{\Delta m^2_{atm}}\right)^{1/3}=\dd\frac{(2|q|^2s)^{1/3}}{h}x\\[10 pt]
m_{\beta\beta}=\dd\frac{m_{0\nu}}{h}\dd\frac{\left\vert x_{23}^2-x_{22}x_{33}\right\vert}{2 h|q|} x~~~.
\end{array}
\end{align}
Making use of eqs. (\ref{expansions}), (\ref{defs}) and (\ref{ofinterest}), we see that the averages of eqs. (\ref{ave}) 
can be reproduced by $x\approx 0.1$, $m_{0\nu}/h=11.5$ meV and
\begin{align} 
\label{approx} 
\begin{array}{l}
\dd\frac{|q|}{h}\approx 2.3~~~~~~~~\dd\frac{s}{h}\approx 3.4~~~~~~~~
\dd\frac{\sqrt{2}|n|}{h}\approx 1.5\\[10pt]
\dd\frac{|l\bar k+\bar l k|}{2\sqrt{2}h s}\approx 1.4~~~~~~~~~
\dd\frac{\left\vert x_{23}^2-x_{22}x_{33}\right\vert}{2 h |q|}\approx 4.8
\end{array}
\end{align}
We conclude that all the dimensionless quantities in eq. (\ref{ave}) scale linearly with $x$,
with proportionality coefficients of order one. The only relatively
large coefficient is that of the combination controlling $m_{\beta\beta}$,
whose value in eq. (\ref{ave}) has the largest relative fluctuation.

This result holds for the most general K\"ahler potential compatible with modular invariance.
If a flavour universal K\"ahler potential is adopted, the anti-holomorphic variable $\bar u$ 
only affects the overall scale of the mass matrices, can be absorbed in the parameter 
$m^0_\nu$ of eq. (\ref{caseb}) and drops from all dimensionless quantities. If so, all the complex
parameters $x_{ij}$ have a common phase, up to a sign. 
The result of eqs. (\ref{expansions}), (\ref{defs}) and (\ref{ofinterest}) is valid for any level $N$
and for all possible assignment of weights $k_\varphi$, provided the combination $k_S$ of eq. (\ref{Omega})
is odd and $m_\nu(0,0)$ is singular. Indeed $k_S$ is odd in all but one of the $14$ pairs of CP invariant models
selected above. If $k_S$ is even, another set of universal scaling laws analogous to those in eqs. (\ref{expansions}) and (\ref{ofinterest}) applies. Their agreement with the data is less satisfactory, since they predict
$\Delta m^2_{sol}/\Delta m^2_{atm}$ of order one and both $\sin\theta_{12}$ and $\sin\theta_{13}$ of order $x$.
A singular $m_\nu(0,0)$ can only arise through the seesaw mechanism. If neutrino masses are described by 
the Weinberg operator and $m_\nu(0,0)$ is regular, we can easily derive other scaling relations predicting both $\Delta m^2_{sol}/\Delta m^2_{atm}$ and $\sin\theta_{13}$ of order $x$, in tension with the data.
Finally, our analysis can be easily extended to the other fixed points.
Under the assumption that $\rho_L$ is irreducible, an expansion around $\tau=\omega$ leads to
universal but phenomenologically disfavored predictions, while a similar analysis around $\tau=i\infty$ 
leads to results depending on the level $N$. The interest in those fixed points is related to the generation
of natural mass hierarchies in the charged lepton sector \cite{Novichkov:2021evw}, an aspect that we have not touched in the present discussion.
\vspace{5pt}

\noindent
{\bf Conclusions}

\noindent
Modular invariance may represent one important piece for the solution of the flavour puzzle. Many concrete bottom-up realizations suggest that a single modulus can dominate the description of the neutrino sector. Nevertheless model building suffers from the large freedom related to the 
choice of the level and the weights, as well as to possible non-universal kinetic terms.
We have shown that, even freely varying all these ingredients, a universal behavior emerges in the vicinity of the
self-dual point $\tau=i$, under the sole assumption that the lepton doublets are assigned to a irreducible representation
of the finite modular group. Few cases survive, one of which perfectly matches the experimental data. If 
the neutrino mass matrix is singular at the fixed point and the discrete parameter $k_S$ is odd, a normal ordering is predicted. The sum of neutrino masses lies around 74 meV. Universal relations characterize the scaling 
of several observable quantities in terms of $|\tau-i|$. The independence of these results from the effective flavour group
$SL(2,\mathbb{Z}_N)$ acting on the matter fields and from the specific form of the kinetic terms is a remarkable
feature of the theory. 
Although universal predictions are highly desirable and constitute an outstanding theoretical property,
there are no reasons a priori for the data to privilege a small region of the modulus around $\tau=i$. 
It is really impressive that, over all the allowed parameter space, this is precisely the region preferred by most of the constructions we have examined.

{\bf Acknowledgements}
%
The author thanks Gui-Jun Ding, Jo\~ao Penedo and Arsenii Titov for very helpful comments and for invaluable assistance during the final revisions of this note. This work was supported by the INFN. 

\newpage

\end{document}